\documentclass[aps,pra,twocolumn,superscriptaddress,groupedaddress]{revtex4}
\usepackage{amssymb}
\usepackage{cancel}
\usepackage{color,graphicx}
\usepackage{amsmath}
\usepackage{amsbsy}
\usepackage{amsthm}
\usepackage{bbm}
\usepackage{epsfig}
\usepackage{lscape}
\usepackage{float}
\usepackage{graphicx}
\usepackage{subfigure}
\usepackage{dcolumn}
\usepackage{bbm}
\usepackage{color,epstopdf}
\usepackage{amscd}
\usepackage{amsfonts}  
\usepackage[]{amsmath}
\usepackage{amssymb}    
\usepackage{mathrsfs}
\usepackage{verbatim}
\usepackage[]{cases}
\usepackage{amsmath}
\usepackage{wasysym}
\usepackage[utf8]{inputenc}
\usepackage[T1]{fontenc}
\usepackage{mathtools}

\newcommand{\Rea}{\mathbb{R}}
\newcommand{\Comp}{\mathbb{C}}
\newcommand{\Id}{\hat{\mathbb{I}}}

\newcommand{\Hi}{\mathcal{H}}

\newcommand{\E}{\mathcal{E}}

\newcommand{\Ex}{\mathbb{E}}

\newcommand{\PS}{(\Omega,\E,P)}

\newcommand{\ket}[1]{| {#1} \rangle}
\newcommand{\bra}[1]{\langle {#1}|}
\newcommand{\braket}[1]{\langle {#1} \rangle}

\makeatletter
\newcommand{\oset}[2]{
   {\mathop{#2}\limits^{\vbox to -.5\ex@{\kern-\tw@\ex@
    \hbox{\scriptsize #1}\vss}}}}
\makeatother

\newcommand\harpr[1]{\mathstrut\mkern2.5mu#1\mkern-11mu\raise1.5ex%
  \hbox{$\scriptscriptstyle\rightharpoonup$}}

\newcommand\harpl[1]{\mathstrut\mkern2.5mu#1\mkern-11mu\raise1.5ex%
  \hbox{$\scriptscriptstyle\leftharpoonup$}}

\begin{document}
\title{The inverse Born problem in contextual probability theories: quantum spin and continuous random variables.}

\author{S. Bacchi}
\email{stefano.bacchi@phd.units.it}
\affiliation{Department of Physics, University of Trieste, Strada costiera 11, 34151 Trieste, Italy}  
\author{L. Curcuraci}
\email{luca.curcuraci@phd.units.it}
\affiliation{Department of Physics, University of Trieste, Strada costiera 11, 34151 Trieste, Italy}
\affiliation{Istituto Nazionale di Fisica Nucleare, Trieste Section, Via Valerio 2, 34127 Trieste, Italy}    
\author{G. Gasbarri}
\email{giulio.gasbarri@ts.infn.it}
\affiliation{Department of Physics, University of Trieste, Strada costiera 11, 34151 Trieste, Italy}
\affiliation{Istituto Nazionale di Fisica Nucleare, Trieste Section, Via Valerio 2, 34127 Trieste, Italy}      
\date{\today}

\begin{abstract}
We revise contextual probability theory and the associated inverse Born problem, solved by the Quantum-Like Representation Algorithm (QLRA) in case of two discrete random variables. After pointing out the special feature and limitations of QLRA for two binary random variables, we generalize the QLRA procedure to solve the inverse Born problem to the case of three binary random variables. 
We analyze the quantum spin model in the light of our results and showed that can be easily obtained from a contextual probability model. 
We furthermore study the inverse Born problem in the case of two continuous random variables, exploiting the general idea underlying QLRA.  

\end{abstract}

\maketitle

\section{Introduction.}\label{sec2}
Quantum mechanics is an amazing theory that is perfectly capable of explaining the results of our current experiments, however it seems that no one deeply understands it.
Its probabilistic structure is very peculiar and seems to not fit with the well understood classical probability theory developed by Kolmogorov.  
During the years several efforts had been made comparing quantum probability with the classical one, and from this comparison some amazing results came out, as the so called  \emph{no-go theorems}, i.e. von Neumann theorem (1932) \cite{von2018mathematical}, Bell theorem (1966) \cite{bell2001einstein}, and  Kochen-Specker theorem (1967) \cite{kochen1975problem}, despite other interesting results having been recently found \cite{pusey2012reality}. 
They put constrains on the possibility of representing a quantum model by means of classical probability and gives a tool to better understand quantum probability.
Among them the Kochen-Specker theorem has very important consequences: when compared with a classical probability model, quantum mechanics turns out to be contextual \cite{sep-kochen-specker}. 
	A single classical probability model is not sufficient to fully describe a quantum system, rather a collection of different probability spaces have to be used.
Such a collection takes the name of a \emph{contextual probability model}~\cite{khrennikov2009contextual} and interestingly enough it can also be obtained from a classical probability model (with a single probability space) by a conditioning procedure~\cite{khrennikov2005interference,khrennikov2005interferenceG}. 
Far more interesting,  a procedure to represent contextual probability models on Hilbert spaces, i.e. providing a solution for the so called inverse Born Problem~\cite{khrennikov2016random},  has been proposed.
This procedure goes under the name of the Quantum Like Representation Algorithm (QLRA) and is typically discussed for a pair of binary (i.e. with two distinguishable outcomes) random variables \cite{khrennikov2009contextual,khrennikov2014ubiquitous,khrennikov2005interference,khrennikov2005interferenceG}.
While some generalizations of the QLRA to ternary (with three possible outcomes) random variables \cite{nyman2011quantum} or more general discrete random variables  \cite{khrennikov2005interferenceG} are available, a discussion of the Hilbert space representation of contextual models with more than two random variables or continuous random variables is still missing.
In this work we generalize the procedure to the case of three binary random variables, obtaining the conditions that contextual probabilities need to satisfy in order to  be  consistently represented on a Hilbert space. Interestingly the constraints that the contextual model should satisfy are more stringent than the ones needed to just represent two binary random variables. Despite this, we show that in the obtained contextual probability model there is still room to accommodate the quantum mechanical spin model. We furthermore propose a generalization of the QLRA procedure to the case of continuous random variables, discussing possible techniques for the estimation of the various quantities needed to complete the procedure. Such generalization turns out to be only a partial solution of the inverse Born problem for two continuous random variables.



\section{Classical probability and Bayes theorem.}\label{sec2.5}

Given a random phenomenon, the classical description of its probabilistic properties is done by the following prescription. First define a set $\Omega$ containing all the elementary events i.e. the possible outcomes of a measurement. Then, define the set $\E$ of all the composite events that one could build from the elements of $\Omega$, which is, from the mathematical point of view, a $\sigma$-algebra over $\Omega$. Finally define a normalized measure $P:\E \rightarrow [0,1]$ assigning to each event in $\E$ a probability of occuring (based on all the available information). A classical probability space is defined as the triple $\PS$. Within this model, the observable quantities characterizing the random phenomenon are the random variables defined on $\Omega$, i.e. the functions $X:\Omega \rightarrow \Rea$. More precisely, in the classical probabilistic model, the set $\mathcal{O}$ of all the possible observable quantities (or simply observables) on the random phenomenon coincides with the set of all the possible random variables on $\Omega$.

A very basic but important fact in classical probability theory is that logical operations involving two events commutes. For example, take two events $F,G \in \E$, which can be always thought as two subsets of $\Omega$. It is well known that the logical conjunction ($\wedge$) and disjunction ($\vee$) of two events can be represented by the intersection $\cap$ and union $\cup$ of the two corresponding sets which are commutative \cite{moretti2013spectral}. In particular $F \cap G = G \cap F$, which at the level of the probability measure implies that $P(F \cap G) = P(G\cap F)$. Similar considerations hold for the union of two events. In classical probability, the conditional probability is \emph{defined} by means of 
\begin{equation}\label{bayesformula}
	P(F|G) := \frac{P(F \cap G)}{P(G)},
\end{equation}
with $P(G)>0$. Assuming also $P(F)>0$, the commutativity of the intersection implies that $P(F|G)P(G) = P(G|F)P(F)$, which is the well known \emph{Bayes theorem}. Suppose we have a collection of sets $\{F_n\}_{n \in I}$ such that $\bigcup_{n \in I} F_n = \Omega$. Clearly $P(F_n|G)P(G) = P(G| F_n)P(F_n)$ holds for any $n \in I$, providing that all the events $F_n$ and the event $G$ have positive probabilities. Summing  over $n$ on both sides, we get
\begin{equation}\label{lawoftotalP}
	P(G) = \sum_{n\in I}P(G| F_n)P(F_n)
\end{equation}
which is the so called \emph{law of total probability}. We emphasize how much this result depends on the commutativity of the intersection  $F_n \cap G = G \cap F_n$, and so on the commutativity of the logical operation on the events. This can be considered as signature of classicality in the probabilistic description.

\section{Contextual probability and the inverse Born problem.}\label{sec3}

We stated in the introduction that  a contextual probability model can be understood as a collection of different probability spaces. 
Here we clarify the statement giving a short review on the basic concepts underlying contextual probability models and we then explain the connection between contextual probability models and the inverse Born problem.

Let us label by $c$ a set of specific conditions that must be realized to verify the specific situation in which a probability measure $P_{c}$ characterizes a random phenomenon. The set of conditions labeled by $c$ is called \emph{context}.
We stress the fact that the nature of the context is not important for the contextual model. The conditions defining a context can be of whatever nature.  For example, in physics a context may be determined by an experimental set up, in social science a context may be the personal history experienced by one or more individuals \cite{khrennikov2014ubiquitous}.
 Label with $\mathcal{C}$ the set of all the possible contexts which can  be realized in the study of a random phenomenon. Let $\mathcal{O}$ be the set of all the observables and $\mathcal{P} := \{P_c^A\}_{c \in \mathcal{C}, A\in \mathcal{O}}$ be a collection of probability measures. A contextual probability model is the triple $(\mathcal{C},\mathcal{O},\mathcal{P})$. Let $A \in \mathcal{O}$ be an observable of a random phenomenon and $\sigma(A)$ the set of all the possible values that $A$ may assume\footnote{In what follows we will use the terms "random variable" and "observable" interchangeably due to the substantial equality of the two notions.  Indeed, for any context $c \in \mathcal{C}$ an observable $A \in \mathcal{O}$ is a random variable on the probability space $(\sigma(A), \E_c^A, P_c^A)$, where $\E_c^A$ is some $\sigma$-algebra over $\sigma(A)$ support of $P_c^A$.}.  For any $A \in \mathcal{O}$ and $P_c^A \in \mathcal{P}$, the quantity $P_c^A(\alpha) \in [0, 1]$ is the \emph{contextual} probability to find $A = \alpha$ when we are in the context $c$. The expectation value of $A$ in the context $c$ is given by the standard formula
\begin{equation*}
	\Ex_c[A] = \sum_{\alpha \in \sigma(A)} \alpha P_c^A(\alpha).
\end{equation*}

Now let $B \in \mathcal{O}$ be another observable and $\sigma(B)$ the set of all the possible values that $B$ may assume.
The conditional probability $P(\beta | \alpha)$ to have $B=\beta$ given $A=\alpha$ can be introduced by using a special family of contexts $\{c(\alpha)\}_{\alpha \in \sigma(A)}$, called \emph{$\alpha$-selection contexts}. Such family of contexts is characterized by the property that $P^A_{c(\alpha)}(\alpha') = \delta_{\alpha,\alpha'}$, which means that, in the context $c(\alpha)$, the probability of finding $A=\alpha$ is one and, consequently, the probability of finding $A\neq\alpha$ is zero. One typically assume that in a contextual probability model, such family of contexts exists for any observable in $\mathcal{O}$. With this in hand one can \emph{define} the conditional probability $P(\beta|\alpha)$ as \cite{khrennikov2009contextual}
\begin{equation}\label{beyondbayes}
	P(\beta|\alpha) := P_{c(\alpha)}^B(\beta).
\end{equation}
We stress the fact that this definition of the conditional probability is not in general equivalent to \eqref{bayesformula}, hence we \emph{do not expect} that Bayes theorem to hold to the general case. An important consequence of this fact is that the law of total probability described by Eq.~\eqref{lawoftotalP} cannot be always applied  for a generic contextual probability model. This crucial feature, distinguishing a contextual probability model from the classical model, is mathematically described by the following modified law of total probability for the observables $A,B \in \mathcal{O}$:
\begin{equation}\label{khrennicovftp}
	P_c^B(\beta) = \sum_{\alpha \in \sigma(A)} P(\beta|\alpha)P_c^A(\alpha) + \delta(\beta|A,c).
\end{equation}
where in the first term on the r.h.s. one recognizes the ordinary law of total probability, while $\delta(\beta|A,c)$  is called \emph{coefficient of $B|A$-supplementarity for the context $c$} and quantify the violation with respect to \eqref{lawoftotalP} in the given contextual probability model for the specific observables and context.

We would like to stress on the fact that, even if contextual probability models seem very different from the classical one, it is always possible to derive a contextual probability model from a classical probability space $\PS$ by conditioning it \cite{khrennikov2005interference,khrennikov2005interferenceG}. Indeed,  given a classical probability space $\PS$ a context can be understood as an event in $\E$, and the $\alpha$-selection context as the event $c(\alpha) = \{A =\alpha\}$.
In this way, the contextual probabilities can be written as ordinary conditional probability $P_c(\cdot) = P(\,\cdot\, |c  )=P(\cdot \cap c)/P(c)$ for any $c \in \mathcal{C} \subset \mathcal{E}$, and $P(\beta|\alpha)$ is given by eq.~\eqref{bayesformula}. However  $P(\beta|\alpha)P_c^A(\alpha) \neq P(\alpha|\beta)P_c^B(\beta)$ and so the standard law of total probability is violated\footnote{Note that in this formulation, the law of total probability still holds between $A$ and $B$ if we use the conditional probabilities $P(\beta|\alpha,c):=P(\{B = \beta\}| \{A=\alpha\} \cap  c)$ for any $c\in\mathcal{C}$. However, the basic assumption of a contextual probabilistic model, when embedded into a classical probabilistic model, is that $P(\beta|\alpha,c)$ are not used for the description of the random phenomenon under study. For this reason one uses \eqref{beyondbayes} to introduce the conditional probabilities. }. To avoid confusion with the conditional/contextual probabilities $P_c$, we refer to $P(\beta|\alpha)$ with the term \emph{transition probability.} 

The possibility to embed a contextual probability model in a "bigger" classical probabilistic model is a known fact (see \cite{holevo2011probabilistic}, Th. 7) however only recently a procedure to represent a specific class of contextual probabilistic models on a Hilbert space has been proposed \cite{khrennikov2005interference,khrennikov2005interferenceG,khrennikov2009contextual}. This procedure goes under the name of \emph{Quantum Like-Representation Algorithm} (QLRA). QLRA  solves the \emph{inverse Born problem}, defined in \cite{khrennikov2016random} as follows:
\begin{center}
	\textquotedblleft \emph{To construct a representation of probabilistic data by complex probability amplitudes that match Born's rule.} \textquotedblright
\end{center}
More precisely, starting from  Eq.~\eqref{khrennicovftp} relating the probability distributions of two observables $A$ and $B$ (which can be derived from some set of probabilistic data), QLRA gives:
\begin{enumerate}
	\item[i)] the vector $|\psi_c\rangle$ on a Hilbert space $\Hi$ such that
	\begin{equation*}
		P_c^A(\alpha) = |\langle \alpha | \psi_c \rangle |^2 \mspace{15mu}\mbox{and}\mspace{15mu} P_c^B(\beta) = |\langle \beta | \psi_c\rangle|^2,
	\end{equation*}
	where $P_c^A(\alpha)$ and $P_c^B(\beta)$ are the contextual probability of $A=\alpha$ and $B=\beta$, respectively, while $\{|\alpha\rangle\}_{\alpha \in \sigma(A)}$ and $\{|\beta\rangle\}_{\beta \in \sigma(B)}$ are two different orthonormal basis of $\Hi$;
	\item[ii)] the representation of the observables $A$ and $B$ on $\Hi$ by means of two operators
	\begin{equation*}
		\hat{A} = \sum_{\alpha \in \sigma(A)} \alpha | \alpha \rangle \langle \alpha | \mspace{15mu}\mbox{and}\mspace{15mu}
		\hat{B} = \sum_{\beta \in \sigma(B)} \beta | \beta \rangle \langle \beta |, 
	\end{equation*}
	such that $\Ex_c[A] = \langle \psi_c | \hat{A} |\psi_c\rangle$ and $\Ex_c[B] = \langle \psi_c| \hat{B}|\psi_c\rangle$. 		  
\end{enumerate}	
From $(i)$ and $(ii)$ one may understand why the QLRA  can be interesting in the field of foundations of quantum mechanics. It is indeed capable to derive key features of the quantum formalism that in the theory must be postulated. 
However QLRA is able to give an ordinary (in the sense of quantum mechanics) Hilbert space representation only for a particular class of contexts, called \emph{trigonometric} (defined in the next section), in which the two observables $A$ and $B$ are symmetrically conditioned~\cite{khrennikov2009contextual,khrennikov2016random},i.e.
\begin{align}\label{symmetricallycond}
	P(\alpha|\beta)=P(\beta|\alpha).
\end{align}
In the next section we briefly introduce the QLRA for binary observables in trigonometric contexts under the assumption of symmetrically conditioned transition probability. The more general case of random variable with $n>2$ different outcomes can be found in \cite{khrennikov2005interferenceG,khrennikov2009contextual,nyman2011quantum}.

\section{QLRA for binary observables.}\label{sec4}

Given a contextual probability model $(\mathcal{C},\mathcal{O},\mathcal{P})$ and two binary random variables $A,B \in \mathcal{O}$, i.e. $\sigma(A) = \{\alpha_1,\alpha_2\}$ and $\sigma(B) = \{\beta_1,\beta_2\}$, equation \eqref{khrennicovftp} can be rewritten as
\begin{align}
	P_c^B(\beta) =& \sum_{\alpha \in \sigma(A)} P(\beta|\alpha)P_c^A(\alpha)\nonumber\\
	 &\hspace{0.5cm}+ 2\lambda(\beta|A,c)\sqrt{\prod_{\alpha \in \sigma(A)}P(\beta|\alpha)P_c^A(\alpha)},
\end{align}
where
\begin{equation}\label{trigonometric}
	\lambda(\beta|A,c) := \frac{\delta(\beta|A,c)}{2\sqrt{\prod_{\alpha \in \sigma(A)}P(\beta|\alpha)P_c^A(\alpha)}}.
\end{equation}
A context $c$ is said to be  \emph{trigonometric} if $|\lambda(\beta|A,c)|\leqslant1$ for all $\beta \in \sigma(B)$. 
Indeed in this case $\lambda(\beta|A,c)$ can be written as a function of a \emph{probabilistic angle} $\theta(\beta|A,c)$ by means of trigonometric functions. In particular,
\begin{equation}\label{probabilisticangle}
	\theta(\beta|A,c) := \arccos(\lambda(\beta|A,c)).
\end{equation}
Note that angle $\theta(\beta |A,c )$ defined by \eqref{probabilisticangle} is well defined if $P_c^A(\alpha)>0$ for all $\alpha \in \sigma(A)$ and, when this happens, the context $c$ is said to be $A$-\emph{non-degenerate}. 
After having introduced the trigonometric contexts we move to resume the QLRA for binary observables.
The QLRA is based on the following identity:
\begin{equation}\label{ordinaryID}
	X^2+Y^2+2XY\cos(\theta) = |X + e^{i\theta}Y|^2. 
\end{equation}
Exploiting the above equation one is able to define the function
\begin{equation}\label{khrennikovwavefunction}
	\psi_c(\beta) := \sqrt{P(\beta|\alpha_1)P_c^A(\alpha_1)} + e^{i\theta(\beta|A,c)}\sqrt{P(\beta|\alpha_2)P_c^A(\alpha_2)},
\end{equation}
which satisfies  $P_c^B(\beta) = |\psi_c(\beta)|^2$ for all $\beta \in \sigma(B)$. 
This function resembles a standard wave function in QM and its structure is completely determined by the probabilistic set of data. However some more steps are needed in order to have the contextual probability model represented in the Hilbert space $\Comp^{2}$.
One first needs to define the vectors
\begin{equation}\label{bbasis}
	|\beta_1\rangle := \begin{pmatrix} 1  \\  0 \end{pmatrix}	\mspace{15mu}\mbox{and}\mspace{15mu}
	|\beta_2\rangle := \begin{pmatrix} 0 \\  1 \end{pmatrix}.
\end{equation}
associated to the observable $B$, which form an orthonormal basis of $\Comp^2$.
Exploiting this basis one can now define the vector in $\Comp^2$ associated to the function \eqref{khrennikovwavefunction} describing the context $c$:
\begin{equation}\label{QLRApsiB}
	|\psi_c\rangle := \sum_{\beta \in \sigma(B)}\psi_c(\beta)|\beta\rangle = \begin{pmatrix} \psi_c(\beta_1) \\  \psi_c(\beta_2) \end{pmatrix}.
\end{equation}
Note that, if $\langle \cdot | \cdot \rangle$ denotes the ordinary scalar product in $\Comp^2$, one has that $P_c^B(\beta) = |\langle \beta | \psi_c \rangle|^2$.
With this in hand is it now easy to see that the contextual expectation value of $B$ can be computed as
\begin{equation*}
	\Ex_c[B] := \sum_{\beta \in \sigma(B)}\beta|\psi_c(\beta)|^2 = \langle\psi_c|\hat{B}|\psi_c \rangle
\end{equation*}
where $\hat{B}$ is an operator acting on $\Comp^2$ defined as
\begin{equation}\label{Boperator}
	\hat{B} = \sum_{\beta \in \sigma(B)} \beta |\beta\rangle\langle \beta|.
\end{equation}
Summarizing, the procedure above returns the vector $|\psi_c\rangle \in \Comp^2$ and the operator $\hat{B}$ that completely characterize the observable $B$ on the Hilbert space $\Comp^2$. What is left is to construct a representation of the observable $A$ in $\Comp^2$ such that (i) and (ii) are satisfied.
Equations \eqref{khrennikovwavefunction} and \eqref{QLRApsiB} suggest that we can define the vectors
\begin{align}\label{abasis}
	|\alpha_1,c\rangle &:= \begin{pmatrix} \sqrt{P(\beta_1|\alpha_1)}  \\  \sqrt{P(\beta_2|\alpha_1)} \end{pmatrix} \hspace{0.5cm}\text{and}\nonumber\\
	|\alpha_2,c\rangle &:= \begin{pmatrix} e^{i\theta(\beta_1|A,c)}\sqrt{P(\beta_1|\alpha_2)} \\  e^{i\theta(\beta_2|A,c)}\sqrt{P(\beta_2|\alpha_2)} \end{pmatrix},
\end{align}
which are associated to the observable $A$ in the basis of the observable $B$. It is important to notice that in this case, on the contrary to \eqref{bbasis},  the vectors $|\alpha,c\rangle$ may depend on the specific context. Using these vectors, \eqref{khrennikovwavefunction} and \eqref{abasis}, one is allowed to rewrite  $|\psi_c\rangle \in \Comp^2$ as
\begin{align}\label{psi-A-basis}
		&|\psi_c\rangle = \begin{pmatrix} \psi_c(\beta_1) \\  \psi_c(\beta_2) \end{pmatrix} \nonumber\\
		&= \begin{pmatrix} \sqrt{P(\beta_1|\alpha_1)P_c^A(\alpha_1)} + e^{i\theta(\beta_1|A,c)}\sqrt{P(\beta_1|\alpha_2)P_c^A(\alpha_2)} \\ \sqrt{P(\beta_2|\alpha_1)P_c^A(\alpha_1)} + e^{i\theta(\beta_2|A,c)}\sqrt{P(\beta_2|\alpha_2)P_c^A(\alpha_2)} \end{pmatrix} \nonumber\\
		&= \!\sqrt{P_c^A(\alpha_1)}|\alpha_1,c\rangle + \sqrt{P_c^A(\alpha_2)}|\alpha_2,c\rangle =:\hspace{-0.2cm} \sum_{\alpha \in \sigma(A)}\hspace{-0.2cm} \psi_c(\alpha)|\alpha,c\rangle. 
\end{align}
Rewriting the vector $|\psi_c\rangle$ in this form suggests that $|\alpha,c\rangle$ are good candidates for the basis of $A$. However, to completely solve the inverse Born problem we have to be sure that (i) and (ii) are satisfied.
From the above expression it is easy to see that the square modulus of the component of $|\psi_c\rangle$ in the direction $|\alpha,c\rangle$ coincides with $P_{c}^{A}(\alpha)$. Consequently (i) and (ii) holds if the vectors in \eqref{abasis} form an orthonormal basis.
In \cite{khrennikov2009contextual} it is proven that orthonormality of the vectors in \eqref{abasis} is guaranteed iff we assume transition probability $P(\alpha|\beta)$ to be symmetrically conditioned. The proof presented there is based on the fact that the matrix 
\begin{equation}\label{changeofbasis}
	\begin{split}
		\hat{U}_{AB} &= \begin{pmatrix} \langle \beta_1|\alpha_1,c\rangle &  \langle \beta_2|\alpha_1,c\rangle \\ \langle \beta_1|\alpha_2,c\rangle & \langle \beta_2|\alpha_2,c\rangle \end{pmatrix} \\
		&=\begin{pmatrix}  \sqrt{P(\beta_1|\alpha_1)} &    \sqrt{P(\beta_2|\alpha_1)}\\
			e^{i\theta(\beta_1|A,c)}\sqrt{P(\beta_1|\alpha_2)} & e^{i\theta(\beta_2|A,c)}\sqrt{P(\beta_2|\alpha_2)}
		\end{pmatrix},
	\end{split}
\end{equation}
mapping the basis $|\beta_{i}\rangle$ into $|\alpha_{i},c\rangle$, is unitary iff the transition probabilities $P(\alpha|\beta)$ are symmetrically conditioned. Indeed the unitarity of $\hat{U}_{AB}$ and orthonormality of $\{|\beta\rangle\}_{\beta \in \sigma(B)}$ are necessary and sufficient conditions for the orthonormality of $\{|\alpha,c\rangle\}_{\alpha \in \sigma(A)}$, since the scalar product is preserved under unitary transformations. 

Summarizing, when condition \eqref{symmetricallycond} holds, the vectors in \eqref{abasis} form an orthonormal basis on  $\Comp^{2}$ , $P_c^A(\alpha) = |\langle \alpha|\psi_c\rangle|^2$ and
\begin{equation*}
	\Ex_c[A] = \sum_{\alpha\in\sigma(A)} \alpha |\psi_c(\alpha)|^2 = \langle \psi_c| \hat{A} |\psi_c \rangle
\end{equation*}
where
\begin{equation}\label{operatorA}
	\hat{A}= \sum_{\alpha \in \sigma(A)} \alpha |\alpha, c\rangle\langle \alpha, c|.
\end{equation}
allowing the representation $A$ on the same $\Comp^2$ in which $B$ is represented.

It is important to stress that if $A$ and $B$ are symmetrically conditioned random variables, the following properties are satisfied \cite{khrennikov2009contextual}:
\begin{enumerate}
	\item[a)] the matrix of transition probabilities $\mathbf{P}(A|B):= [ P(\beta|\alpha) ]_{\beta \in \sigma(B),\alpha \in\sigma(A)}$ is doubly stochastic, that in the case of binary random variables reduces to  $P(\beta|\alpha)=1/2$.
	\item[b)] the two phases $\theta(\beta_1|A,c)$ and $\theta(\beta_2|A,c)$ are such that
	\begin{equation}\label{phasecondition}
		e^{i\theta(\beta_1|A,c)} = - e^{i\theta(\beta_2|A,c)};
	\end{equation}
	\item[c)] the dependence of $|\alpha,c\rangle$ on the context $c$ is removed, i.e. $|\alpha,c\rangle = |\alpha\rangle$.
\end{enumerate}
Condition (c) is very important because guarantees that, once the Hilbert space is fixed by the observable $B$, it is possible to represent the observable $A$ in the same Hilbert space and its representation will not depend on the context c.
Instead conditions (a) and (b) allow us to simplify eq.~\eqref{abasis}. Indeed, 
exploiting (a) and (b) and  calling  $\theta= \theta(\beta_1|A,c)$, the $A$ basis in Eq.~\eqref{abasis} can be rewritten as
\begin{equation}\label{abasis2}
	|\alpha_{1}\rangle = \frac{1}{\sqrt{2}}\begin{pmatrix}1  \\ 1  \end{pmatrix}	\mspace{15mu}\mbox{and}\mspace{15mu}
	|\alpha_2\rangle = \frac{e^{i\theta}}{\sqrt{2}}\begin{pmatrix} 1 \\  -1 \end{pmatrix}.
\end{equation}
Consequently the change of basis matrix $\hat{U}_{AB}$ takes form
\begin{equation}\label{VAB}
	\hat{U}_{AB} = \frac{1}{\sqrt{2}}\begin{pmatrix} 1 & 1 \\ e^{i\theta} & -e^{i\theta} \end{pmatrix},
\end{equation}
which can be used to show that the diagonal matrix $\hat{A}$, representing the observable $A$, is not in general diagonal in the basis of $B$. Indeed, 
\begin{equation}\label{Aoperator}
	\hat{A} = \hat{U}_{AB}^{\dagger} \begin{pmatrix} \alpha_1 & 0 \\ 0 & \alpha_2 \end{pmatrix}  \hat{U}_{AB} =\frac{1}{2}\begin{pmatrix} \alpha_1+\alpha_2 & \alpha_1 - \alpha_2 \\ \alpha_1 - \alpha_2 &  \alpha_1+\alpha_2 \end{pmatrix}.
\end{equation}
This is an important fact because shows the non commutativity of the two observables, i.e. $[\hat{A},\hat{B}] \neq0$, outlining the non-classicality of the contextual probabilistic model.

Equation~\eqref{Aoperator} also shows that the matrix representing A in the basis of B must be real, suggesting that QLRA applied on two binary random variables is not able to reproduce operators with complex entries. We also note that the state $|\psi_c\rangle$ has only real components in the $|\alpha\rangle$-basis. These two features characterize the QLRA, and are consequences of definition \eqref{abasis2} given for the $|\alpha\rangle$-basis, as observed in \cite{khrennikov2009contextual}. 
One may be tempted to modify definition \eqref{abasis2} to remove this restriction of the algorithm, however any modification of QLRA  will not allow the Hilbert space representation of the probabilistic model to be completely specified by the probabilistic data set.
For example one could perform the transformation $\theta\to \theta- \omega$ in the $\ket{\alpha}$-basis to obtain a new orthonormal basis,
\begin{equation}\label{newabasis}
	|\alpha_1\rangle ' = |\alpha_1\rangle  \mspace{15mu} \mbox{ and } \mspace{15mu} |\alpha_2\rangle' = e^{-i\omega}|\alpha_2\rangle.
\end{equation}
and a new  matrix of the change of basis, i.e.
\begin{equation}\label{VABome}
	\hat{U}_{AB}^{\omega} =  \frac{1}{\sqrt{2}}\begin{pmatrix} 1 & 1 \\ e^{i(\theta-\omega)} & -e^{i(\theta-\omega) }\end{pmatrix}.
\end{equation}
It is not difficult to verify that the representation of the operator $\hat{A}$ in the $|\beta\rangle$-basis is not changed by the transformation. However the representation of the vector \eqref{psi-A-basis} is changed. In particular $|\psi_{c}\rangle$ is given by,
\begin{equation}\label{Awave}
	|\psi_c\rangle =  \sqrt{P_c^A(\alpha_1)}|\alpha_1\rangle' + \sqrt{P_c^A(\alpha_2)}e^{i\omega}|\alpha_2\rangle'.
\end{equation}
Showing that in the new basis \eqref{newabasis} the $|\psi_{c}\rangle$ vector can have complex amplitudes. However, one immediately notices that the freedom of the choice of $\omega$ is completely arbitrary, meaning that this degree of freedom is irrelevant for the probabilistic model under consideration. We will show in the next sections how, the introduction of a third observable can lead to observables represented by operators with complex entries.



\section{QLRA for three binary observables.}\label{sec5b}

Suppose we have a contextual probability model $(\mathcal{C},\mathcal{O},\mathcal{P})$ with three binary observables $A,B,C \in \mathcal{O}$, pairwise symmetrically conditioned with $\sigma(A)=\{\alpha_{1},\alpha_{2}\}$, $\sigma(B)=\{\beta_{1},\beta_{2}\}$ and, $\sigma(C)=\{\gamma_{1},\gamma_{2}\}$. We can apply the standard QLRA procedure to all different pairs of the three observables.
\\ 

We first apply QLRA to the pair $(B,A)$ to obtain the wave vector $\ket{\psi_{c}}$ representing the context c, the $\ket{{\alpha}}$ and $\ket{\beta}$ basis (see previous section) of the Hilbert space $\Comp^{2}$, the change of basis matrix $\hat{U}_{AB}$ and the operators $\hat{A}=\sum_{i}\alpha_{i}\ket{\alpha_{i}}\!\bra{\alpha_{i}}$ and $\hat{B}=\sum_{i}\beta_{i}\ket{\beta_{i}}\!\bra{\beta_{i}}$  representing   the observables A and B on $\Comp^2$.
We then apply QLRA to $(C,A)$ pair\footnote{Note that with the writing $(C,A)$ we emphasize that $C$ plays the same role of the observable $B$ for the couple $(B,A)$.} to obtain a new wave vector $\ket{\tilde{\psi}_{c}}$, two new basis $|\tilde{\alpha}\rangle$ and $|\gamma\rangle$ connected by the matrix of the change of basis $\hat{U}_{AC}$(see eq.~\eqref{VAB} where  the $\beta$'s are replaced by the $\gamma$'s and $\theta$ by $\phi = \phi(\gamma_1 | A,c)$), and the operators $\hat{A'}=\sum_{i}\alpha_{i}\ket{\tilde{\alpha}_{i}}\!\bra{\tilde{\alpha}_{i}}$ and $\hat{C}=\sum_{i}\gamma_{i}\ket{\gamma_{i}}\!\bra{\gamma_{i}}$ representing the observables  $A$ and $C$ on the Hilbert space $\Comp^2$.
We notice that the representation of the wave vector $\ket{\psi}_{c}$ in the $\ket{\alpha}$-basis, i.e.
\begin{align}\label{eq:alpha}
	\ket{\psi_{c}}= \sqrt{P_c^A(\alpha_1)}|\alpha_1\rangle + \sqrt{P_c^A(\alpha_2)}|\alpha_2\rangle =: \sum_{\alpha \in \sigma(A)} \psi_c(\alpha)|\alpha\rangle,
\end{align}
and the representation of the wave vector $\ket{\tilde{\psi}_c}$ in the $\ket{\tilde{\alpha}}$-basis, i.e.
\begin{align}
	\ket{\tilde{\psi}_{c}}= \sqrt{P_c^A(\alpha_1)}|\tilde{\alpha}_1\rangle + \sqrt{P_c^A(\alpha_2)}|\tilde{\alpha}_2\rangle =: \sum_{\alpha \in \sigma(A)} \psi_c(\alpha)|\tilde{\alpha}\rangle,
\end{align}
have exactly the same structure~\footnote{This fact is not obvious in general. Indeed choosing e.g the pair (A,C)  instead of (C,A) will not allow for the same result.}. Because of this we may assume
\begin{align}
	\ket{\alpha_{1}}=\ket{\tilde{\alpha}_{1}} \hspace{0.5cm} \text{and}\hspace{0.5cm} \ket{\alpha_{2}}=\ket{\tilde{\alpha}_{2}},
\end{align}
implying $\ket{\psi_{c}}=|\tilde{\psi}_{c}\rangle$. 
Under this assumption we are allowed to construct the transformation matrix ($\hat{W}_{CB}$) connecting $|\beta\rangle$-basis and $|\gamma\rangle$-basis, as the product of the $\hat{U}_{AC}^{\dagger}$  time the $\hat{U}_{AB}$  matrix, i.e.
\begin{equation}\label{generalchangeofbasis}
	\hat{W}_{CB} = [\hat{U}_{AC}]^\dagger\hat{U}_{AB} =\begin{pmatrix} w_1 & w_2 \\w_2 & w_1 \end{pmatrix}
\end{equation}
where
\begin{align}\label{equaw0}
	w_1 &:= \frac{1 + e^{i(\theta-\phi)}}{2} , \hspace{0.5cm}
	w_2 := \frac{1-e^{i(\theta-\phi)}}{2}
\end{align}
and, from that, obtain 
\begin{equation}\label{csystem}
	\begin{cases}
		|\gamma_1\rangle &= w_1|\beta_1\rangle + w_2|\beta_2\rangle \\
		|\gamma_2\rangle &= w_{2}|\beta_1\rangle + w_1|\beta_2\rangle.
	\end{cases}
\end{equation}

Interestingly the coefficients that connect the $|\gamma\rangle$-basis to the $|\beta\rangle$-basis are functions of the probabilistic angle $\theta$ describing the pair $(B,A)$ and the probabilistic angle $\phi$ describing the pair $(C,A)$.
This fact imposes constraints on the probabilistic model that allows a representation of three observables in Hilbert space through QLRA as we see next.
Let us calculate $P(\beta|\gamma)=|\braket{\beta|\gamma}|^{2}$. With the help of Eq.~\eqref{csystem} we obtain
\begin{align}\label{eq:errr}
	P(\beta_{1}|\gamma_{1})=P(\beta_{2}|\gamma_{2})= |w_{1}|^{2} = \cos^{2}((\theta-\phi)/2)\nonumber\\
	P(\beta_{1}|\gamma_{2})=P(\beta_{2}|\gamma_{1})= |w_{2}|^{2} = \sin^{2}((\theta-\phi)/2)
\end{align}
showing that the matrix $P(\beta|\gamma)$ must be symmetrically conditioned and consequently doubly stochastic.
Double stochasticity for binary random variables implies that $P(\beta|\gamma)=1/2$, that replaced in eq.~\eqref{eq:errr}  gives
\begin{align}
	\cos^{2}\left(\frac{\theta-\phi}{2}\right)=\sin^{2}\left(\frac{\theta-\phi}{2}\right)=\frac{1}{2},
\end{align}
condition satisfied iff
\begin{align}\label{eq:tfconf}
	\theta(\beta_1|A,c) - \phi(\gamma_1|A,c) = \frac{\pi}{2} \mod 2\pi.
\end{align}

Let us turn our attention on the state vector $\ket{\psi_{c}}$. 
The QLRA applied to the pair $(C,A)$ leads to	$|\psi_c\rangle = \sum_{i}\psi_{c}(\gamma_{i})\ket{\gamma_{i}}$
with

\begin{align}\label{eq:psigamma}
	\psi_{c}(\gamma_{1})&= \frac{\sqrt{P_c^A(\alpha_1)} + \sqrt{P_c^A(\alpha_2)}e^{i\phi}}{\sqrt{2}},\nonumber\\
	\psi_{c}(\gamma_{2})&= \frac{\sqrt{P_c^A(\alpha_1)} - \sqrt{P_c^A(\alpha_2)}e^{i\phi}}{\sqrt{2}}.
\end{align}

Exploiting  eq.~\eqref{csystem} we can rewrite the state-vector in $|\beta\rangle$-basis, i.e. $|\psi_c\rangle = \sum_{i}\psi_{c}(\beta_{i})\ket{\beta_{i}}$ and obtain
\begin{align}\label{consistency-cond}
	\psi_c(\beta_1) =  w_1^*\psi_c(\gamma_1) + w_2^{*}\psi_c(\gamma_2), \nonumber\\
	\psi_c(\beta_2) = w_2^*\psi_c(\gamma_1) + w_1^{*}\psi_c(\gamma_2).
\end{align}
Performing the square modulus of the first equation in \eqref{consistency-cond} and making use of $P_{c}(\beta)=|\psi_{c}(\beta)|^{2}$ and $P(\beta|\gamma)=|\braket{\beta|\gamma}|^{2}$ we obtain:
\begin{align}\label{temporary}
	&P_c^B(\beta_1)  = | w_1^*\psi_c(\gamma_1) + w_2^{*}\psi_c(\gamma_2) |^2 \nonumber\\
	&=\sum_{i=1}^{2}P(\beta_1|\gamma_i)P_c^C(\gamma_i)+ 2\Re \left[ w_{1}w_{2}^{*}\psi_c(\gamma_1)^*\psi_c(\gamma_2) \right]\nonumber\\
	&=  \sum_{i=1}^{2}P(\beta_1|\gamma_i)P_c^C(\gamma_i) + \sqrt{P_c^A(\alpha_1)P_c^A(\alpha_2)} \sin( \phi )
\end{align}
where for the last line we used eqs.~\eqref{equaw0},~\eqref{eq:psigamma} and \eqref{eq:tfconf}.
We then write the contextual law of total probability  for the pair (B,C) i.e. 
\begin{align}\label{BC-LTP}
	P_c^B(\beta_1) &= \sum_{\gamma \in \sigma(C)} P(\beta_1|\gamma)P_c^C(\gamma)\nonumber\\
	& + 2\cos(\chi(\beta_1|C,c))\sqrt{\prod_{\gamma \in \sigma(C)}P(\beta_1|\gamma)P_c^C(\gamma)}.
\end{align}
and by comparison between eq.~\eqref{temporary} and eq.~\eqref{BC-LTP} we obtain:
\begin{equation}\label{consistency-condition}
	\begin{split}
		\cos(\chi(\beta_1|C,c)) &= \sqrt{\frac{P_c^A(\alpha_1)P_c^A(\alpha_2)}{P_c^C(\gamma_1)P_c^C(\gamma_2)}} \sin(\phi(\gamma_1|A,c)) \\
		&=  \sqrt{\frac{P_c^A(\alpha_1)P_c^A(\alpha_2)}{P_c^C(\gamma_1)P_c^C(\gamma_2)}} \cos(\theta(\beta_1|A,c)),
	\end{split}
\end{equation}
where on the last step we used the relation \eqref{eq:tfconf} among the probabilistic angles and $P(\beta_1|\gamma_1) = P(\beta_1|\gamma_2) = 1/2$.  Essentially equivalent condition can be obtained starting from the second equation of \eqref{eq:psigamma}.
Equation \eqref{consistency-condition} gives a condition that probabilistic angles and contextual probabilities must satisfy in order to allow a consistent representation of the three binary observables in the same Hilbert space.
To conclude this section is interesting to rewrite the matrix representing the operator $\hat{C}$ in the $\ket{\beta}$-basis.
With help of $\hat{W}_{CB}$ we obtain
\begin{equation}\label{Coperator}
	\begin{split}
		\hat{C} &= \hat{W}_{CB}^\dagger\begin{pmatrix} \gamma_1 & 0 \\ 0 & \gamma_2 \end{pmatrix}  \hat{W}_{CB}  \\
		&= \begin{pmatrix} |w_1|^2\gamma_1 + |w_2|^2\gamma_2& (\gamma_1-\gamma_2) w_1^*w_2 \\ (\gamma_1 - \gamma_2)w_1w_2^* & |w_2|^2\gamma_1 + |w_1|^2\gamma_2 \end{pmatrix}.
	\end{split}
\end{equation}
This equation shows that the operator $\hat{C}$ in $\beta$-basis is in general represented by a complex matrix.
It is important to observe that the entries of the matrix in eq.~\eqref{Coperator} can be fully determined by the probabilistic data at our disposal, meaning that the complexity of the matrix is a consequence of some specific feature of the probabilistic model. We stress that this procedure cannot be equivalent to apply the QLRA directly to the couple (C,B) because, the standard QLRA procedure do not allow for complex entries in the operator C when represented in the $\ket{\beta}$-basis, meaning that, in the case of more than two observables, we cannot solve the inverse Born problem just "naively" applying the QLRA.

The procedure proposed in this section shows that a contextual probabilistic model allows for a complex representation of the observables in Hilbert space, when more than two observables are present.
It also shows that a representation of three observables can be obtained by iteratively applying QLRA among two of the three pairs of the observables.
However the procedure does not allow a representation of generic contextual probability model.
Indeed, for the procedure to work, we need the contexts $c$ to be trigonometric for any possible pairing of observables and observables to be pairwise symmetrically conditioned, as in the standard QLRA for two observables, but we  also need the contextual probabilities to satisfy  the relation in eq.~\eqref{consistency-condition}.
Even if  all of these conditions seems very restrictive, we show in the next section that the class of contextual probability models allowing for a consistent representation of three observables in Hilbert space is  general enough to reproduce quantum spin, an important model in the field of quantum theory.

\subsection{Example: the quantum spin.}\label{sec6}
In this section we show how the procedure developed in the previous section is able to reproduce quantum spin.
We will show that the QLRA for three observables developed in the previous section is able to reproduce the Pauli matrices and furthermore allows for a generic spin state $\ket{\psi_{c}}$ describing the context.

Suppose that $ \theta - \phi = - \pi/2$ and the eigenvalues of $A$, $B$ and $C$  are respectively
\begin{equation}\label{eq:asspaul}
	\beta_1 = -\beta_2 = 1, \mspace{15mu} \alpha_1 = - \alpha_2 = 1, \mspace{15mu} \gamma_1 = - \gamma_2 = 1.
\end{equation}
Using \eqref{Boperator}, \eqref{Aoperator}, \eqref{Coperator} and  \eqref{equaw0} one  obtains
\begin{equation*}
	\hat{B} =  \begin{pmatrix} 1 & 0 \\ 0 & -1\end{pmatrix}, \mspace{15mu} \hat{A} =  \begin{pmatrix} 0 & 1 \\ 1 & 0\end{pmatrix}, \mspace{15mu} \hat{C} =  \begin{pmatrix} 0 & i \\ -i & 0\end{pmatrix}.
\end{equation*}
that respectively are $\sigma_{z}$, $\sigma_{x}$ and $\sigma_{y}$ Pauli matrices.
What is left to check is if the state $\ket{\psi_{c}}$, describing the context, is represented by a generic state and not only by a restricted class of states. 
We  recall that given a basis $\{ |v_1\rangle,|v_2\rangle\}$ the most general state for a two-level system takes the following form
\begin{equation*}
	|\psi\rangle = G |v_1\rangle + \sqrt{1-G^2}e^{iF}|v_2\rangle,
\end{equation*}
where $G \in [0,1]$ and $F \in [0,2\pi]$, up to an overall phase. 
The representation  $\ket{\psi_{c}}$ of the context in the $\ket{\beta}$ and $\ket{\gamma}$-basis can be evidently written in this form,
but $\ket{\psi_{c}}$ is purely real if  represented in $\ket{\alpha}$-basis, as one can see from eq.~\eqref{eq:alpha}. 
This fact seems to suggest that the context is described by only the wave vectors that are purely real in the $\ket{\alpha}$-basis. 
However, this problem can be easily overcome applying the transformation in \eqref{eq:alpha} in both of the pairs $(B,A)$ and $(C,A)$ described in the previous section, i.e. $\theta\to \theta-\omega$ and $\phi\to \phi-\omega$, with the price of a free parameter $\omega$ to be fixed.\footnote{
	We stress that this procedure does not change any of the results of the previous section because the matrix of the change of basis $\hat{W}_{CB}$  depends on the difference $\theta-\phi$.}
A reasonable choice to fix the free parameter is $\omega=\theta$. 
Indeed under this assumption one obtains a representation of the $\ket{\alpha}$-basis independent of the contextual probabilities $P_{c}^{A}$.
Looking at \eqref{abasis2}, we can see that the presence of $\theta$ implies that the $|\alpha\rangle$-basis depends on the contextual probabilities $P_c^A$~\footnote{Hence for each state $|\psi_c\rangle$ of the random phenomenon, the $|\alpha\rangle$-basis changes.}, but after the transformation in \eqref{eq:alpha} and assuming $\omega=\theta$ the new $\ket{\alpha}$-basis becomes:
\begin{align}
	|\alpha_{1}\rangle = \frac{1}{\sqrt{2}}\begin{pmatrix}1  \\ 1  \end{pmatrix}	\mspace{15mu}\mbox{and}\mspace{15mu}
	|\alpha_2\rangle =  \frac{1}{\sqrt{2}}\begin{pmatrix} 1 \\  -1 \end{pmatrix}.
\end{align}
The equation not only shows that the $\ket{\alpha}$-basis is now independent of the contextual probability $P_{c}^{A}$ but it is also the standard representation of the $\sigma_{x}$ eigenvectors.
It is easy to check from eq.~\eqref{bbasis} that the $\ket{\beta}$-basis also gives the standard representation of the $\sigma_{z}$ eigenvectors, but the $\ket{\gamma}$-basis does not give the standard representation of the $\sigma_{y}$ eigenvectors.
Indeed from \eqref{equaw0} and \eqref{csystem} with $\theta-\phi =-\pi/2$ we obtain
\begin{equation*}
	|\gamma_1\rangle = \zeta|y_+\rangle, \mspace{30mu}  |\gamma_2\rangle =\zeta^*|y_-\rangle,
\end{equation*}
with $\zeta := e^{-i\frac{\pi}{4}}$ and 
\begin{equation*}
	|y_+\rangle = \frac{1}{\sqrt{2}}\begin{pmatrix}
		1 \\ i
	\end{pmatrix},
	\mspace{30mu}
	|y_-\rangle = \frac{1}{\sqrt{2}}\begin{pmatrix}
		1 \\ -i
	\end{pmatrix},
\end{equation*}
are the $\sigma_{y}$ eigenvectors in the standard representation.
Hence $|\gamma_1\rangle$ and $|\gamma_2\rangle$ are equal up to a phase factor to the standard eigenvectors of $\hat{\sigma}_y$.  The presence of this phase factor does not seem to alter the probabilistic description of the spin model, however its origin is not completely understood by the authors and remains a curious fact that needs to be investigated in the future.

\section{QLRA for continuous random variables}

In this section we attempt to extend QLRA to the case of continuous random variables. 
We work under the assumption that the probability distribution involved admits density with respect to the Lebesgue measure. 
As we will see, we do not solve the inverse Born problem stated in section \ref{sec3}, rather a \emph{weaker} version of it. By this we mean that we can only provide a solution of the point $ii)$ of the inverse Born problem, while the wave vector can be determined only up to a local phase factor.

We start by writing the continuous version of the law of total probability \eqref{khrennicovftp}. Given a contextual probability space $(\mathcal{O},\mathcal{C},\mathcal{P})$, two continuous observables $A,B \in \mathcal{O}$ and a context $c\in\mathcal{C}$, the continuous law of total probability can be written as
\begin{equation}\label{continuousFTP}
	\rho_c^B(b) = \int p(a|b)\rho_c^A(a)da + \omega(b|A,c).
\end{equation}
Wthere $p(b|a)$, $\rho_c^B(b)$ and $\rho_c^A(a)$ are respectively the transition probabilities density and the contextual probability densities of $B$ and $A$ in the context $c$ and $\omega(b|A,c)$ measures the violation of the (continuous variant of) Bayes theorem in the contextual probability model and we call it \emph{supplementarity density}. 
We aim to define a vector $|\psi_c\rangle$ in some Hilbert space $\Hi$ and two generalized basis $\{|\alpha\rangle\}$ and $\{|\beta\rangle\}$ such that:
\begin{equation}\label{Continuousrequirement}
	|\langle a | \psi_c\rangle|^2 = \rho_c^A(a),\mspace{15mu} |\langle b|\psi_c\rangle|^2=\rho_c^B(b)\, ,\mspace{15mu} |\braket{a|b}|^{2}=p(a|b) 
\end{equation}
and 
\begin{align}\label{generalizedbasis}
	\begin{split}
		\int \ket{a}\!\bra{a}d a &=\int \ket{b}\!\bra{b}d b  =\Id_{\Hi} \\
		\langle a|b\rangle &= \langle b|a\rangle^*, \\
		\langle a|a'\rangle &= \delta(a-a'), \\
		\langle b|b'\rangle &= \delta(b-b').
	\end{split}
\end{align}
The requirement $|\braket{a|b}|^{2}=p(a|b)$ combined with $\braket{a|b}=\braket{b|a}^{*}$ force us to assume $A$ and $B$ to be symmetrically conditioned, i.e. $p(b|a) = p(a|b)$. Indeed from these two conditions we can immediately obtain $p(a|b)=|\braket{a|b}|^{2}=|\braket{b|a}|^{2}=p(b|a)$.
Calling $\psi_{c}(a)=\braket{a|\psi_{c}}$ we can write:
\begin{align}\label{generalisedpsi_a}
	|\psi_c\rangle = \int \psi_c(a)|a\rangle da = \int |\psi_c(a)|e^{i\xi_c(a)}|a\rangle da
\end{align}
where $\xi_c(a)$ is a phase function that must be determined through the available statistical data.
Under the assumptions that $\rho_{c}^{B}(b)=|\braket{b|\psi_{c}}|^{2}$ and $\{\ket{a}\}$ is a generalized basis fulfilling \eqref{generalizedbasis}, we can write:
\begin{equation*}
	\begin{split}
		\rho_c^B(b) &= \langle \psi|b\rangle\langle b|\psi\rangle \\
		&= \int da\int da' \langle \psi|a\rangle\langle a|b\rangle\langle b|a'\rangle\langle a'|\psi\rangle \\
		&= \int da\int da' |\psi_c(a)||\psi_c(a')|\\
		&\hspace{1cm}\times\sqrt{p(a|b)p(a'|b)}e^{i[\xi_c(a')-\xi_c(a) + \eta(a,b) - \eta(a',b)]}
	\end{split}
\end{equation*}
where in the last line we have used the fact that $\braket{a|b}=\sqrt{p(a|b)}e^{i\eta(a,b)}$ where $\eta(a,b)$ is a phase function that should be fixed by statistical data.
Exploiting \eqref{Continuousrequirement} and the symmetry of   $\sqrt{p(a|b)p(a'|b)\rho_c^A(a)\rho_c^A(a')}$,  we can rearrange  this equation as:
\begin{align}\label{continuousFTPHil}
	&\rho_c^B(b)= \int\!\! da  p(b|a)\rho_c^A(a) \nonumber\\
	&+\!\! \int\!\! da\!\! \int\!\! da' [1\! -\! \delta(a\!-\!a'\!)]\sqrt{\!p(a|b)p(a'|b)\rho_c^A(a)\rho_c^A(a')}e^{i\theta_b\!(a,a'\!|c)}
\end{align}
with
\begin{equation}\label{localphase}
	\theta_b(a,a'|c) := \xi_c(a')-\xi_c(a) + \eta(a,b) - \eta(a',b).
\end{equation}
Comparing 
Eq.~\eqref{continuousFTPHil} with Eq.~\eqref{continuousFTP} we obtain
\begin{align}\label{Delta}
	&\omega(b|A,c) = \nonumber\\
	&\int\!\! da\! \int\!\! da' [1\! -\! \delta(a-a')]\sqrt{\!p(a|b)p(a'|b)\rho_c^A(a)\rho_c^A(a')}e^{i\theta_b(\!a,a'\!|c)}.
\end{align}
If we now assume that
\begin{align}\label{continuoustrigonometric}
	\frac{1}{2\sqrt{p(a|b)p(a'|b)}}\frac{\delta}{\delta \sqrt{\rho_c^A(a)}}\frac{\delta}{\delta\sqrt{\rho_c^A(a')}} \omega(b|A,c) \in [-1,1].
\end{align}
we can reverse Eq.~\eqref{Delta} and obtain an explicit expression for $\theta_b(a,a'|c)$ in function of the statistical data of our disposal (see Appendix A for further details), i.e. 
\begin{align}\label{angoloBAc}
&	\theta_b(a,a'|c)=\nonumber\\
&\begin{cases}
		0 &\mbox{if}\mspace{5mu} a = a' \\
		\!\arccos\! \left(\!\! \frac{1}{2\sqrt{\!p(\!a|b)p(\!a'\!|b)}}\frac{\delta}{\!\delta\! \sqrt{\!\rho_c^A(a)}}\!\frac{\delta}{\delta\!\sqrt{\!\rho_c^A(a')}}\omega(b|A,\!c)\!\! \right) &\mbox{if}\mspace{5mu} a \neq a'
	\end{cases}
\end{align}
Note that eq.~\eqref{continuoustrigonometric} is the continuous equivalent of the trigonometric context condition \eqref{trigonometric} used in the case of binary and discrete random variables.
However eq.~\eqref{angoloBAc} is not enough to solve the inverse Born problem as defined in section \ref{sec3}. It does not allow all the free parameters of the model to be fixed from the the probabilistic data.
Nevertheless we can still solve a weaker version of the inverse Born problem under the further assumption that the wave-function $\ket{\psi_c}$ is real in the $\ket{\alpha}$-basis, i.e. 
\begin{equation}\label{eq:areal}
	|\psi_c\rangle = \int \sqrt{\rho_c^A(a)}|a\rangle da,
\end{equation}
which means that $\xi_c(a)=0$ for all $a \in \sigma(A)$. Note that this assumption corresponds to eq.~\eqref{psi-A-basis} in the binary case.
This assumption simplifies eq.~\eqref{angoloBAc} as
\begin{align}
	\theta_b(a,a'|c) :=\eta(a,b) - \eta(a',b),
\end{align}
and the observable $B$, whose representing operator is diagonal in the $|b\rangle$-basis, can be represented in $|a\rangle$-basis as
\begin{equation*}
	\begin{split}
		\hat{B} &:= \int b |b\rangle \langle b| db = \int \int  \left(\int db b \langle a| b\rangle \langle b| a'\rangle\right) |a\rangle \langle a'| dada' \\
		&= \int \int  \beta(a,a') |a\rangle \langle a'|dada',
	\end{split}
\end{equation*}
where the matrix elements $\beta(a,a')$ are given by
\begin{align*}
	\beta(a,a') &= \int b \sqrt{p(a|b)p(a'|b)}e^{i(\eta(a,b)-\eta(a',b)}db\nonumber\\
	 &= \int b \sqrt{p(a|b)p(a'|b)}e^{i\theta_b(a,a')}db.
\end{align*}
As one can see these matrix elements can be computed using \eqref{angoloBAc} and are completely determined by the statistical data of our probabilistic model. 
Exploiting the representation of $\ket{\psi_{c}}$ and $\hat{B}$ in the $\ket{a}$-basis we can then compute the expectation values of $B$. Similarly we can compute the matrix elements and expectation value of $\hat{A}$ representing the observable $A$. 
The proposed continuous generalization of QLRA is not able to produce a fixed representation of the contextual probability model in the Hilbert space. Indeed, inserting an identity in \eqref{generalisedpsi_a}, the wave vector in the $\ket{b}$-basis is
\begin{equation*}
	\ket{\psi_c} = \int \psi_c(b)|b\rangle db
\end{equation*}
with
\begin{equation*}
	\psi_c(b) = \int da \sqrt{p(a|b)\rho_c(a)}e^{i\eta(a,b)}.
\end{equation*}
But the method here proposed is only able to relate the difference $\eta(a,b) - \eta(a',b)$ to the supplementarity density $\omega(b|A,c)$ which can be estimated from the probabilistic data. Meaning that the wave vector $\ket{\psi_c}$ cannot be completely determined by probabilistic data with the proposed procedure. However, this representation can still be used to compute the expectation values of the $A$ and $B$ observables. 
Even if this procedure solves a $\emph{weak}$ version of the inverse Born problem, eq.~\eqref{angoloBAc} is quite cumbersome and it is not generally clear how to perform functional derivative over the the supplementarity density. In the next section we attempt to solve this problem, using the introduction of new observables.

\subsection{The role of additional random variables}

The beauty of the ordinary QLRA is the simplicity with which is able to construct a quantum like representation  from  the statistical data \cite{khrennikov2014ubiquitous}. However, the continuous generalization of the QLRA here presented does not seem to share the same property. In particular it seems rather cumbersome to estimate the functional dependence (crucial for the computation of \eqref{angoloBAc}) of the supplementarity density from the statistical data, . 

A strategy for the estimation of \eqref{angoloBAc}  may be the following.
By definition we have
\begin{equation*}
	\omega(b|A,c) = \rho_c^B(b) - \int p(b|\tilde{a})\rho_c^A(\tilde{a})d\tilde{a},
\end{equation*}
from which we obtain the fact that the functional derivative with respect to $\sqrt{\rho_c^A(a)}$ and $\sqrt{\rho_c^A(a')}$  with $a\neq a'$ over $\omega(b|A,c)$ can be written as 
\begin{equation*}
	\frac{\delta}{\delta \sqrt{\rho_c^A(a)}}\frac{\delta}{\delta\sqrt{\rho_c^A(a')}} \omega(b|A,c) = \frac{\delta}{\delta \sqrt{\rho_c^A(a)}}\frac{\delta}{\delta\sqrt{\rho_c^A(a')}}\rho_c^B(b)
\end{equation*}
Thus the problem of computing $\eqref{angoloBAc}$ reduces to the computation of the second functional derivative of $\rho_c^B(b)$. 
In order to compute the r.h.s of this equation we can consider two more observables, $X,Y \in \mathcal{O}$, which are related to $A$ and $B$ through the law of total probability \eqref{continuousFTP}. Exploiting the chain rule for functional derivatives we rewrite equation above as 
\begin{align}\label{chainrule}
	&\frac{\delta}{\delta \sqrt{\rho_c^A(a)}}\frac{\delta}{\delta\sqrt{\rho_c^A(a')}} \omega(b|A,c) \nonumber\\
	&=\!\! \int\!\! dx\!\! \int\! dy \frac{\delta \sqrt{\rho_c^X(x)}}{\delta \sqrt{\rho_c^A(a)}}\frac{\delta \sqrt{ \rho_c^Y(y)}}{\delta\sqrt{\rho_c^A(a')}}\frac{\delta}{\delta\sqrt{\rho_c^X(x)}}\frac{\delta}{\delta\sqrt{\rho_c^Y(y)}}\rho_c^B\!(b).
\end{align}
The r.h.s term of this equation  can now be computed by exploiting the law of total probabilities that relates $A$, $B$, $X$ and $Y$. 
Exploiting the law of total probability that relates $X$ to $A$ we obtain
\begin{equation*}
	\frac{\delta \sqrt{\rho_c^X(x)}}{\delta \sqrt{\rho_c^A(a)}} = \frac{1}{2 \sqrt{\rho_c^X(x)}}\left[ p(x|a)\sqrt{\rho_c^A(a)} + \frac{\delta \omega(x|A,c)}{\delta \sqrt{\rho_c^A(a)}} \right]
\end{equation*}
and a similar result holds also for $\frac{\delta \sqrt{ \rho_c^Y(y)}}{\delta\sqrt{\rho_c^A(a')}}$.
To compute the second functional derivative of $\rho_c^B(b)$ in eq.~\eqref{chainrule}, we  proceed as follows.
We write $\rho_c^B(b) = (\sqrt{\rho_c^B(b)})^2$, we then use the law of total probability associated to the observable $X$ to rewrite one of  the $\sqrt{\rho_c^B(b)}$ and the law of total probability associated to the observable $Y$ to rewrite the other  $\sqrt{\rho_c^B(b)}$ on the equation, to obtain
\begin{widetext}
\begin{equation}\label{eq:pirop}
	\begin{split}
		&\frac{\delta}{\delta\sqrt{\rho_c^X(x)}}\frac{\delta}{\delta\sqrt{\rho_c^Y(y)}}\rho_c^B(b) = \frac{1}{4\rho_c^B(b)}\left[ p(b|x)\sqrt{\rho_c^X(x)} + \frac{\delta \omega(b|X,c)}{\delta \sqrt{\rho_c^B(b)}} \right]\left[ p(b|y)\sqrt{\rho_c^Y(y)} + \frac{\delta \omega(b|Y,c)}{\delta \sqrt{\rho_c^Y(y)}} \right]  	
	\end{split}
\end{equation}
\end{widetext}
We now have to compute the functional derivative of the other four supplementarity densities, in order to do so we can iterate the method.
Iterating the method will produce an infinite series and it is not clear if it can be truncated at a certain step.
Another possibility is that we could assume that the two observables $X$ and $Y$ are such that the functional derivatives that appear in eq.~\eqref{eq:pirop} are identically zero (this would happen, for instance, if the context $c$ is a classical-like context \cite{khrennikov2009contextual} for these additional observables). 
Maybe considerations on mutually unbiassed basis and contextuality in general can help to find a strategy to find such observables that allow the explicit solution of eq.~\eqref{eq:pirop}, however we couldn't find it.

Another way to solve the problem could be to embed the contextual probability model into a bigger Kolmogorov  probability space~\footnote {This can be always done as observed in \cite{holevo2011probabilistic,khrennikov2005interference,khrennikov2005interferenceG}.} i.e. assuming explicit knowledge of the \emph{contextual transition probabilities} between the observables (see section \ref{sec2.5}). 
Taken two auxiliary observables  $X,Y \in \mathcal{O}$, one may write
\begin{align}\label{eq:ltpd}
	\rho_c^B(b) &= \int p(b|x,c)\rho_c^X(x)dx \mspace{15mu}\mbox{and}\nonumber\\
	 \rho_c^A(a)& = \int p(a|x,c)\rho_c^X(x)dx
\end{align}
and the same for $Y$, where $p(b|x,c)$ are the  \emph{contextual transition probability densities}, i.e. the densities of $P(\{B \in [b_1,b_2]\}|\{X \in [x_1,x_2]\}\cap c)$ with $b_1,b_2 \in \sigma(B)$ and $x_1,x_2 \in \sigma(X)$. 
Exploiting eq.~\eqref{eq:ltpd}, and the corresponding for $Y$, it is easy to obtain 
\begin{equation*}
	\begin{split}
		\frac{\delta \sqrt{\rho_c^X(x)}}{\delta \sqrt{\rho_c^A(a)}} &= \frac{p_c(x|a)\sqrt{\rho_c^A(a)}}{2 \sqrt{\rho_c^X(x)}} \\
		\frac{\delta \sqrt{\rho_c^Y(y)}}{\delta \sqrt{\rho_c^A(a')}} &= \frac{p_c(y|a')\sqrt{\rho_c^A(a')}}{2 \sqrt{\rho_c^Y(y)}} \\
		\frac{\delta}{\delta\sqrt{\rho_c^X(x)}}\frac{\delta}{\delta\sqrt{\rho_c^Y(y)}}\rho_c^B(b) &= \frac{p_c(b|x)p_c(b|y)\sqrt{\rho_c^X(x)\rho_c^Y(y)} }{4\rho_c^B(b)} 
	\end{split}
\end{equation*}
which can be used in \eqref{chainrule} to explicitly compute the second functional derivative.
Although this method is able to give us a closed formula for the probabilistic angle, it assumes knowledge of the contextual transition probabilities, something that is not usual in a contextual probability model. 
However this procedure may be useful in the derivation of a contextual probability model  from underlying Kolmorogovian models.

Summarizing, the continuous version of QLRA does not share the nice easy-to-use properties of the ordinary QLRA. However, it could be a useful tool in the investigation of the role of observables in a contextual probability model, and may have relevant application in the study of contextual probability models as restrictions of bigger Kolmogorovian models.

\section{Conclusion}

In this work we generalized the QLRA procedure to solve the inverse Born problem for the case of three binary random variables symmetrically conditioned in the trigonometric contexts, and we proposed a generalization of the QLRA  able to solve a weakened version of the inverse Born problem for the case of two continuous symmetrically conditioned random variables.
We showed that it is possible to construct an Hilbert space representation for three observables of a contextual probability model by properly iterating the ordinary QLRA procedure developed for two binary observables at the price of additional constraints involving both the contextual probability and the probabilistic angles. \\
We showed the emergence of the complex structure of the operators (peculiar to quantum mechanics) when more than two observables are considered\footnote{In the three observable case the operators representing the observables must be hermitians but can have complex entries, in contrast with the case in which only two observables are considered, where the operators must be purely real.}.
We analyzed the quantum spin model in the light of our results and showed that it can be easily obtained from a contextual probability model once that the possible outcomes of the observables are specified.\\
These are very interesting facts which suggest the possibility of understanding quantum probability as a contextual probability model. Regarding continuous random variables, it seems that only a weak version of the inverse Born problem can be solved following the line of reasoning of ordinary QLRA. However, this solution is satisfying if one is only interested in reproducing the expectation values and the probabilities of the theory. This may be considered as a first partial solution of the continuous version of the inverse Born problem that needs further investigation.
\section*{Acknowledgments}

LC and GG acknowledge C. Jones for her thoughtful comments. 
LC acknowledges A. Ferrari for many useful discussion on this topic.
GG would also like to thank the European Union's Horizon 2020 research and innovation programme under grant agreement No 766900 [TEQ] for founding support.

\bibliographystyle{unsrt}
\bibliography{biblioqlra}

\onecolumngrid

\appendix

\section{}

In this section we provide a proof of eq.~\eqref{angoloBAc}.
From eq.~\eqref{localphase} one immediately see that $\theta_b(a,a'|c) = 0 $ for $a = a'$.
Now consider $a \neq a'$. 
We rewrite eq.~\eqref{Delta} as:
\begin{align}
	&\omega(b|A,c) = \int_{a\neq a' } da da' \sqrt{\!p(a|b)p(a'|b)\rho_c^A(a)\rho_c^A(a')}e^{i\theta_b(\!a,a'\!|c)} 
\end{align}
and we take the first functional derivative of eq.~\eqref{continuoustrigonometric}, to obtain: 
\begin{equation*}
	\begin{split}
		&\frac{\delta}{\delta \sqrt{\rho_c^A(a)}} \omega(b|A,c) = \int_{\tilde{a}\neq \tilde{a}'} d\tilde{a} d\tilde{a}' \sqrt{p(\tilde{a}|b)p(\tilde{a}'|b)}e^{i\theta_b(\tilde{a},\tilde{a}'|c)} \frac{\delta}{\delta \sqrt{\rho_c^A(a)}} \sqrt{\rho_c^A(\tilde{a})\rho_c^A(\tilde{a}')} \\
		&= \int_{\tilde{a}\neq \tilde{a}'} d\tilde{a} d\tilde{a}' \sqrt{p(\tilde{a}|b)p(\tilde{a}'|b)}e^{i\theta_b(\tilde{a},\tilde{a}'|c)} \left[ \delta(a-\tilde{a})\sqrt{\rho_c^A(\tilde{a}')} + \sqrt{\rho_c^A(\tilde{a})}\delta(a - \tilde{a}')  \right]\\
		&= \int d\tilde{a}' \sqrt{p(a|b)p(\tilde{a}'|b)}e^{i\theta_b(a,\tilde{a}'|c)}\sqrt{\rho_c^A(\tilde{a}')} + \int d\tilde{a} \sqrt{p(\tilde{a}|b)p(a|b)}e^{i\theta_b(\tilde{a},a|c)}\sqrt{\rho_c^A(\tilde{a})}
	\end{split}
\end{equation*}
we then take the second functional derivative with $a'\neq a $ to get:
\begin{equation*}
	\begin{split}
		\frac{\delta}{\delta \sqrt{\rho_c^A(a')}} \frac{\delta}{\delta \sqrt{\rho_c^A(a)}} \omega(b|A,c) = \sqrt{p(a|b)p(a'|b)}\left[ e^{i\theta_b(a,a'|c)} + e^{i\theta_b(a',a|c)} \right]
	\end{split}
\end{equation*}
Now we notice that $\theta_b(a,a'|c) = - \theta_b(a',a|c)$ (See eq.~\eqref{localphase}) and rewrite the above equation as 
\begin{equation*}
	\frac{\delta}{\delta \sqrt{\rho_c^A(a')}} \frac{\delta}{\delta \sqrt{\rho_c^A(a)}} \omega(b|A,c) = 2\sqrt{p(a|b)p(a'|b)}\cos \theta_b(a,a'|c),
\end{equation*}
from which eq.~\eqref{angoloBAc} follows.  Note that the condition \eqref{continuoustrigonometric} ensures that the cosine is well define.

\end{document}